\documentclass[12pt]{iopart}

\usepackage{iopams}  
\usepackage{graphicx}

\begin{document}

\title[Thermometry with spin-dependent lattices]{Thermometry with spin-dependent lattices}

\author{D. McKay and B. DeMarco}

\address{Department of Physics, University of Illinois, 1110 W Green St., Urbana, Il 61801}
\ead{dcmckay2@illinois.edu}
\begin{abstract}
We propose a method for measuring the temperature of strongly correlated phases of ultracold atom gases confined in spin-dependent optical lattices. In this technique, a small number of ``impurity'' atoms---trapped in a state that does not experience the lattice potential---are in thermal contact with atoms bound to the lattice. The impurity serves as a thermometer for the system because its temperature can be straightforwardly measured using time-of-flight expansion velocity. This technique may be useful for resolving many open questions regarding thermalization in these isolated systems. We discuss the theory behind this method and demonstrate proof-of-principle experiments, including the first realization of a 3D spin-dependent lattice in the strongly correlated regime.
\end{abstract}

\maketitle

\section{Introduction}

Ultracold atoms trapped in optical lattices are an ideal system for probing strongly correlated quantum phases. Recent results in these systems include the observation of a fermionic Mott insulator \cite{schneider:2008,jordens:2008}, superexchange in a Bose-Hubbard system \cite{trotzky:2008}, a disordered strongly interacting insulator \cite{white:2009b}, and in-situ imaging of a 2D bosonic Mott insulator \cite{gemelke:2009}. While one of the main goals remains observing antiferromagnetism \cite{koetsier:2008}, current cooling techniques remain insufficient to reach the low entropies required to obtain magnetic ordering. There have been a number of new cooling techniques proposed \cite{bernier:2009,ho:2009}, but measuring temperature remains an outstanding problem if these proposals are to be realized and validated. In addition, measuring temperature is necessary for quantum simulation \cite{zhou:2009}---temperature may be a primary axis of a phase diagram of interest, and a lack of thermodynamic information can lead to ambiguity about observed phases \cite{diener:2007,pollet:2008}. \\

Temperature can be indirectly measured in optical lattice experiments by assuming adiabaticity and equating the entropy in the lattice to the entropy before loading into the lattice \cite{jordens:2008,trotzky:2009}. The initial entropy is straightforward to determine because the lattice is loaded from a weakly interacting gas in a harmonic trap, which is well understood thermodynamically.  Thermometry is then performed by functionally relating entropy in the lattice to temperature. However, in most cases, calculating this relationship is computationally intensive or impossible (i.e., if the physics is unknown).  Also, non-adiabatic heating processes, such as spontaneous emission, cause entropy to be generated in the lattice \cite{trotzky:2009}. And, there has been a recent result \cite{hung:2009} showing that adiabaticity may be difficult to maintain while turning on the lattice.\\

To circumvent these limitations, there have been a number of direct thermometry methods proposed and realized including measuring site-occupancy statistics \cite{kohl:2006}, in-situ diameter \cite{mckay:2009}, spin separation in a two-component Mott insulator \cite{weld:2009}, in-situ number fluctuations \cite{gemelke:2009,zhou:2009}, and direct comparison of time-of-flight images to quantum Monte Carlo simulations \cite{trotzky:2009}.  A general feature of these methods is that they measure a specific aspect of the system under study that has a known relationship with temperature in certain limits.  Therefore, there is some restriction to their applicability since a specialized measurement apparatus and extensive computations may be required, and a reliable theory is necessary.  Most of these approaches are therefore of limited usefulness for optical lattice quantum simulation, which ultimately must probe unknown physics in an unbiased manner.\\

A more general approach is to build an ideal thermometer, which is a system with an exactly understood dependence of measurable quantities on temperature in thermal contact with the system under study. The presence of the thermometer must be non-perturbative, so that the behaviour of the system of interest is unaffected. In this paper, we propose a method to realize such a thermometer---a weakly interacting, harmonically confined gas in thermal contact with strongly correlated lattice atoms---and present proof-of-principle experiments.\\

There are two techniques to prevent the thermometer atoms from experiencing the optical lattice potential.  The first is to use two distinct atomic species, which encounter different optical potentials for the same laser wavelength due to dissimilar electronic structure \cite{leblanc:2007}.  At a specific wavelength the potential can vanish for one species, which has been recently used in a K-Rb mixture to demonstrate a 1D species-specific lattice \cite{catani:2009}.  We pursue another implementation: a spin-dependent potential, where the system and thermometer are in two different internal ``spin'' (i.e. hyperfine) states of the same atomic species.  In this scheme, the lattice potentials are made dependent on the hyperfine state of the atoms by manipulating the laser wavelength and polarization. There are technical advantages to this method since only one atomic species is required. Spin-dependent potentials have applicability beyond thermometry, as they may be used to observe exotic phases (see \cite{zapata:2009}, for example), study four-wave mixing of matter waves \cite{pertot:2009}, and to study thermalization in isolated quantum systems, which is an open question theoretically \cite{rigol:2008} and experimentally \cite{kinoshita:2006}. Also, spin impurites have been studied in a number of contexts, i.e. \cite{palzer:2009,chikkatur:2000,schirotzek:2009}.\\

In the following, we discuss creating spin-dependent optical lattices appropriate for this type of thermometry, and we present experimental results on creating a thermalized impurity and on loading spin mixtures into a 1D and 3D spin-dependent lattice. We also present the first demonstration of atoms trapped in a 3D spin-dependent lattice in the strongly correlated regime. This paper is organized as follows.  Section \ref{sect:theory1} examines the theory of a spin-dependent lattice.  Section \ref{sect:theory2} discusses the theory of co-trapped harmonically confined and lattice-bound gases.  Section \ref{sect:impurity} discusses creating an impurity spin to act as the thermometer; we also present the observation of dynamical ``melting'' of an impurity condensate far from equilibrium.  Section \ref{sect:1Dlattice} presents our implementation of a 1D spin-dependent lattice and evidence that the lattice can exchange energy with atoms in a spin-sensitive fashion.  Finally, Section \ref{sect:3dlattice} will present evidence for a SF-MI transition of a two-component mixture in a 3D spin-dependent lattice.  We also show preliminary results on co-trapping a strongly correlated lattice-bound and weakly interacting harmonically confined gas.

\section{\label{sect:theory1} Spin-Dependent Lattices}

Optical fields can be used to create atomic potentials because neutral atoms interact with an oscillating electric field through the electric dipole interaction. For a simple two-level system, the AC Stark shift of the electronic ground state due to this interaction is
\begin{equation}
V(r) = - \frac{3\pi c^2 \Gamma}{2\omega_0^3} \left(\frac{1}{\omega_0-\omega} + \frac{1}{\omega_0+\omega}\right) I(r), \label{eqn:shift1}
\end{equation}
where $I$ is the laser intensity, $c$ is the speed of light, $\omega_0$ is the atomic transition frequency, $\omega$ is the laser frequency, and $\Gamma$ is the decay rate of the excited state. We have assumed that the detuning $\Delta = \omega-\omega_0>>\Gamma$ and that $I/I_{sat} << 1$, so that the population in the excited state is small. This energy shift can be either positive or negative depending on whether the laser frequency is larger (``blue-detuning'') or smaller (``red-detuning'') than the atomic transition frequency.  For the remainder of this paper, we ignore the small contribution of the counter-rotating term (second inside the brackets in (\ref{eqn:shift1})) given the relatively small detuning required for realizing spin-dependent lattices.  Since the energy shift is proportional to the intensity, the AC Stark shift can be used to confine atoms as the field intensity can have local minima or maxima. Furthermore, by interfering two beams of the same frequency at an angle $\theta$, an optical lattice potential $V_0 \sin^2\left(\frac{k_{lat}x}{\sin(\theta/2)}\right)$ can be created with periodicity $\lambda/2\sin(\theta/2)$, where $k_{lat}$ is the lattice wavevector \cite{schneider:2008,jordens:2008,trotzky:2008,gemelke:2009}.\\

\begin{figure}
\begin{center}
\includegraphics[width=4in]{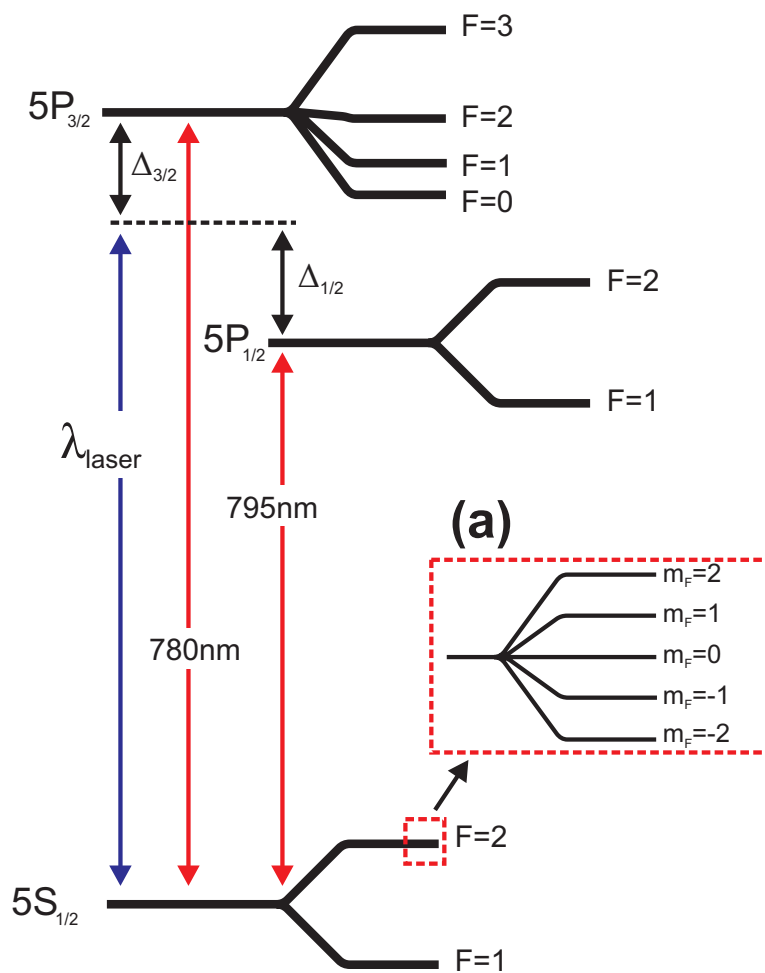}
\end{center}
\caption{Level structure of $^{87}$Rb (not to scale). The blue line represents the applied laser field with detunings $\Delta_{3/2}$ and $\Delta_{1/2}$ from the D2 and D1 transitions, respectively. Inset (a) shows the Zeeman states for each hyperfine level, which are split in a magnetic field with first-order energy shift $g_F m_F B$. Each hyperfine state has $2F+1$ Zeeman states. \label{fig:rbstates}}
\end{figure}

In real atoms, the two-level approximation is not accurate because there are a number of excited state levels.  For example, the level structure of $^{87}$Rb, shown in Figure \ref{fig:rbstates}, has 24 states in the (first) excited-state 5P manifold.  When the laser frequency detuning is large compared to the Zeeman and hyperfine splittings of the excited states, calculating the AC Stark shift by summing over these states is an excellent approximation, and (\ref{eqn:shift1}) becomes \cite{grimm:2000,deutsch:1998},
\begin{eqnarray}
\fl V(r) = \frac{\pi c^2 \Gamma}{2\omega_0^3} \left[ \left(\frac{2}{\Delta_{3/2}}+\frac{1}{\Delta_{1/2}}\right)I(r) + g_F m_F \sum_{q=-1,0,1} q \left(\frac{1}{\Delta_{3/2}}-\frac{1}{\Delta_{1/2}}\right) I_q(r)\right], \label{eqn:statedeplightshift}
\end{eqnarray}
where $m_F$ is the Zeeman state (with gyromagnetic ratio $g_F$) of the atom; $\Delta_{3/2}$ ($\Delta_{1/2}$) is the detuning $\omega-\omega_0$ relative to the $S\rightarrow P_{3/2}$ ($S\rightarrow P_{1/2}$) transition; $q$ refers to the three possible polarizations of light, which are defined with respect to the quantizing magnetic field along the z-axis
\begin{equation}
\hat{\pi} = \left(\begin{array}{c}0\\0\\1\end{array}\right), \hat{\sigma}^{+} = \frac{1}{\sqrt{2}}\left(\begin{array}{l}1\\i\\0\end{array}\right), \hat{\sigma}^{-} = \frac{1}{\sqrt{2}}\left(\begin{array}{c}1\\-i\\0\end{array}\right); 
\end{equation}
and $I_q$ is the intensity of the light with polarization $q$ ($I(r)=\sum_{q=-1,0,1}I_q (r)$). We write (\ref{eqn:statedeplightshift}) assuming that $\Gamma/\omega_0^3$ is the same for the D1 and D2 transitions, which is an excellent approximation for the alkali atoms. The first term in brackets in (\ref{eqn:statedeplightshift}) is the scalar light shift, which is the same for all Zeeman states.  Creating a spin-dependent lattice relies on the tensor shift (the second term in brackets) \cite{deutsch:1998}, which is non-zero only for $q\ne0$ and $g_F m_F \ne0$.\\

\begin{figure}
\begin{center}
\includegraphics[width=4in]{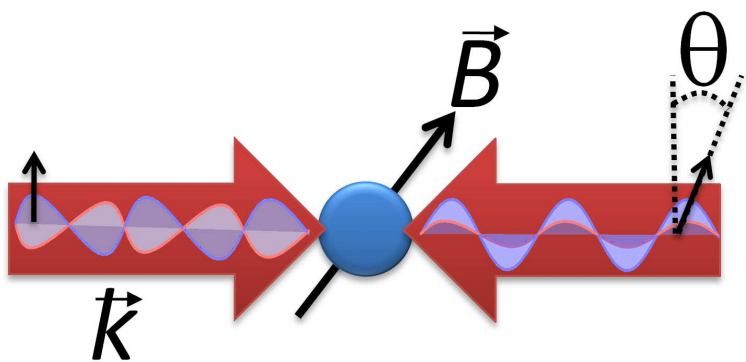}
\end{center}
\caption{Spin-dependent lattice geometry.  The atom gas is shown as a blue circle.\label{fig:linthetalin}}
\end{figure}

To calculate spin-dependent lattice potentials, we adopt a geometry in which the magnetic field is given by the vector $\vec{B}$ ($\hat{B}=\vec{B}/|\vec{B}|$) and the linearly polarized lattice laser beam has a wavevector $\vec{k}$ ($\hat{k}=\vec{k}/|\vec{k}|$). The beam is retro-reflected (with wavevector $-\vec{k}$), and the retro-reflected polarization is rotated with respect to the incoming polarization by an angle $\theta$. This scenario is illustrated in Figure \ref{fig:linthetalin} and is referred to as a lin-$\theta$-lin lattice.  Experimentally, the polarization rotation can be accomplished using a quarter-wave plate or an electro-optical modulator if dynamic polarization rotation is desired. This scenario has been previously considered as a quantum information science tool \cite{jaksch:1999,brennen:1999,franzosi:2006} and has been used experimentally to realize controlled collisions \cite{mandel:2003,mandel:2003b}, quantum walks \cite{karski:2009}, and (with some elaboration) a 2D lattice of double wells \cite{lee:2007}. Previous work on lin-$\theta$-lin lattices, however, only treated the magnetic field and lattice laser wavevector as collinear. In this work, we are concerned with the general case, for which
\begin{eqnarray}
\fl V(r) = \frac{I_0 \pi c^2 \Gamma}{\omega_0^3} \left\{ \left(\frac{2}{\Delta_{3/2}}+\frac{1}{\Delta_{1/2}}\right)\left[1+\cos(\theta)\cos(2\vec{k}\cdot\vec{r})\right] + \right. \nonumber \\
 \left. g_F m_F  \left(\frac{1}{\Delta_{3/2}}-\frac{1}{\Delta_{1/2}}\right) (\hat{k}\cdot\hat{B}) \sin(\theta) \sin(2\vec{k}\cdot \vec{r})\right\}, 
\end{eqnarray}
where $I_0$ is the intensity of the lattice laser beam. This can be re-written as a single sinusoidal potential
\begin{eqnarray}
\fl V(r) = A + \sqrt{A^2\cos^2(\theta) + C^2 (g_F m_F)^2 (\hat{k}\cdot\hat{B})^2 \sin^2(\theta)} \nonumber \\
\cos\left[2\vec{k}\cdot\vec{r} - \textrm{atan} \left(\frac{Cg_Fm_F (\hat{k} \cdot \hat{B})}{A } \tan(\theta) \right) \right], \label{eqn:statedeplattice}
\end{eqnarray}
with
\begin{eqnarray}
A & = & \frac{I_0 \pi c^2 \Gamma}{\omega_0^3} \left(\frac{2}{\Delta_{3/2}}+\frac{1}{\Delta_{1/2}}\right), \textrm{ and}\\
C & = & \frac{I_0 \pi c^2 \Gamma}{\omega_0^3} \left(\frac{1}{\Delta_{3/2}}-\frac{1}{\Delta_{1/2}}\right).
\end{eqnarray}

The basis for the proposed thermometry method is a lin-perp-lin lattice ($\theta=90^{\circ}$), for which the potential is
\begin{equation}
V(r) = A + C(g_F m_F)(\hat{k}\cdot\hat{B})\sin\left(2\vec{k}\cdot\vec{r}\right). \label{eqn:statedeplattice2}
\end{equation}
In this configuration there is no lattice potential for states with $g_F m_F=0$. Our proposal is to use a gas with atoms in the $m_F=0$ state as a thermometer; these atoms are co-trapped with $m_F\ne0$ atoms using a far-detuned, state-independent dipole trap.  In principle, the lattice-bound atoms can then be used to explore strongly-correlated phases while a small number of $m_F=0$ atoms remain weakly interacting and in thermal contact with the gas of interest.  Because the $m_F=0$  atoms are trapped harmonically and are low density, straightforward time-of-flight expansion velocity can be used to determine their temperature.  Bosonic atoms are required for the thermometer, since only they possess $m_F=0$ states.  For a multi-dimensional lattice the magnetic field must be selected so that the $\hat{k}\cdot\hat{B}$ factor is non-zero along all lattice directions.\\

A lin-$\theta$-lin lattice has other features which may be useful for exploring interacting spin physics \cite{ripoll:2004}.  When $\theta=90^{\circ}$, the potentials for states with opposite signs of $g_F m_F$ are $180^{\circ}$ out of phase, i.e., the minimum of the lattice for one state is a maximum for the other.  The lattice potential depth is also proportional to $g_F m_F$, so that co-trapped spin states can experience significantly different lattice potentials.  Both the offset between potential minima and the relative lattice depths are tunable by adjusting $\theta$ (\ref{eqn:statedeplattice}).  Changing $\theta$ can therefore be used to tune the inter- and intra-species interaction strength as well as the relative tunnelling energies \cite{white:2009}.\\

An important consideration for optical dipole potentials is heating caused by momentum diffusion \cite{gordon:1980}.  Even when the AC Stark shift vanishes (e.g., for $m_F=0$ and $\theta=90^{\circ}$), there can still be heating. The rate of energy (``heating power'') increase due to momentum diffusion for a two level atom is
\begin{equation}
\frac{dE}{dt} = \frac{\Gamma}{2m} \left(\frac{|\vec{\mu}_{ge}\cdot \vec{E}|^2}{\Delta^2}\right) \left(k^2 + \left|\frac{\nabla(\vec{\mu}_{ge}\cdot \vec{E})}{\vec{\mu}_{ge}\cdot \vec{E}}\right|^2 \right),
\end{equation}
where $\vec{E}$ is the electric field, $\vec{\mu}_{ge}=\langle g|\vec{\mu}|e \rangle$ is the dipole matrix element and $m$ is the atomic mass. We have omitted counter-rotating terms by assuming $|\Delta| << \omega_0$. For multi-level atoms we sum over all the excited-state levels, and the full heating power is,
\begin{equation}
\fl \dot{E} = E_R \sum_{q} \frac{\pi c^2 \Gamma^2}{2\hbar \omega_0^3} I_q(r) \left(1 + \frac{1}{k^2}\left|\frac{\nabla \vec{E}_q(r)}{\vec{E}_q(r)}\right|^2\right) \left(\frac{2-qm_Fg_F}{\Delta_{3/2}^2} + \frac{1+qm_Fg_F}{\Delta_{1/2}^2}\right),
\end{equation}
where $E_R$ is the recoil energy, and we have assumed $\Gamma_{3/2}^2/\omega_{3/2}^3 \approx \Gamma_{1/2}^2/\omega_{1/2}^3$, which is correct for Rb to within 5\%. For a retro-reflected lin-$\theta$-lin lattice where the forward beam intensity is $I_0$ the heating power is,
\begin{equation}
\dot{E} = E_R \frac{2 \pi c^2 \Gamma^2}{\hbar \omega_0^3} I_0 \left(\frac{2}{\Delta_{3/2}^2} + \frac{1}{\Delta_{1/2}^2}\right),
\end{equation}
which is independent of the angle ($\theta$), the $m_F$ state, the projection of the wavevector on the magnetic field ($\hat{B}\cdot \hat{k}$), and \emph{position in the lattice}. The independence of the heating power on position in the lattice is counter-intuitive for standing wave potentials ($\theta=0$), since one would naively expect the heating rate to vanish at the nodes where the light intensity is zero.  Although heating induced by recoil from scattering is absent in this case, interactions of the fluctuating atomic dipole with the electric field gradient (maximal at the nodes) still results in heating. In fact, heating from dipole fluctuations at the nodes is equal to recoil heating at the anti-nodes, as pointed out in \cite{gordon:1980}.\\

\begin{table}
\caption{\label{depthtable} Comparison of lattice depths ($s$), heating power ($\dot{E}$), and $\tau=s/\dot{E}$ at different wavelengths for $I_0=50W/cm^2$  (approximately 5 mW of light focused to a 80 $\mu$m waist) in the forward lattice beam for a retro-reflected configuration. We assume that $\vec{k}$ and $\vec{B}$ are parallel.}
\begin{indented}
\item[]\begin{tabular}{@{}llllll}
\br
& &\centre{2}{$\theta=0^{\circ}$}&\centre{2}{$\theta=90^{\circ}$}\\
& & & &\centre{2}{$g_Fm_F=1/2$}\\
& & \crule{2} & \crule{2} \\
$\lambda$(nm) & $\dot{E}(E_R/\mathrm{sec})$ & $s$($E_R$) & $\tau(\mathrm{sec})$ & $s$($E_R$) & $\tau(\mathrm{sec})$ \\
\mr
765 & 0.27 & 15.31 & 57.02 & 1.47 & 5.46\\
775 & 2.13 & 42.69 & 19.58 & 6.74 & 3.17\\
785 & 2.86 & 31.88 & 11.16 & 15.62 & 5.47\\
788 & 1.59 & 11.36 & 7.16 & 13.90 & 8.76\\
790 & 1.83 & 0.12 & 0.06 & 15.63 & 8.52\\
805 & 0.42 & 19.72 & 47.48 & 3.32 & 7.99\\
815 & 0.13 & 12.16 & 91.64 & 1.24 & 9.36\\
\br
\end{tabular}
\end{indented}
\end{table}

In Table \ref{depthtable} we compare the lattice potential depth $s$ and scattering rate at $\theta=0^{\circ}$ and $\theta=90^{\circ}$ for states with $g_F m_F=1/2$ and for several lattice laser wavelengths.  The ratio of $\theta=90^{\circ}$ to $\theta=0^{\circ}$ lattice depth is maximized at $\lambda=790$~nm, which we find useful for minimizing complications introduced by imperfect laser polarization (see Section \ref{sect:impurity}).  Furthermore, the ratio $\tau=s/\dot{E}$ is nearly maximum at 790~nm, which also makes this wavelength optimal for realizing spin-dependent lattices.  A disadvantage of the spin-dependent lattice is that the maximum $\tau$ in this range is approximately 9sec, whereas for the  $\theta=0^{\circ}$ lattice $\tau$ is about 50sec for $\lambda=805$~nm.  Unlike the  $\theta=0^{\circ}$ lattice, which can have an arbitrarily high $\tau$ (for arbitrarily high detunings), $\tau$ for a spin-dependent lattice is roughly independent of detuning.  This problem can be alleviated by using atoms with a larger fine structure splitting, such as Cs. 

\section{\label{sect:theory2} Harmonic Gas and Lattice Gas}

In this section we address several practical issues relevant to realizing the proposed thermometry technique. To calculate questions relevant to thermodynamics and interactions between the atoms, we consider a gas of interacting lattice bosons co-trapped with a gas of harmonically confined bosons using the grand canonical ensemble.  If the lattice atoms are labeled with the subscript $\alpha$ and the harmonically trapped atoms with the subscript $\beta$, then the grand canonical Hamiltonian can be written as

\begin{eqnarray}
\fl \hat{K} = -J\sum_{<i,j>}(\hat{a}^{\dagger}_{i,\alpha}\hat{a}_{j,\alpha}) + \sum_{i}\left[\frac{U}{2} \hat{n}_{i,\alpha} (\hat{n}_{i,\alpha}-1) + \left(\frac{1}{2}m\omega_{\alpha}^2 d^2 i^2 -\mu_{\alpha}\right)\hat{n}_{i,\alpha}\right] + \nonumber \\
\sum_{k} \left[\hbar \omega_{\beta} (k+1/2)-\mu_{\beta}\right] \hat{n}_{k,\beta} + \gamma_{\alpha \beta} \int d^3r \hat{n}_{\beta}(r) \sum_{i} \hat{n}_{i,\alpha} \phi^2_{i}(r)
\end{eqnarray}
in the tight-binding limit and neglecting interactions between the $\beta$ atoms.  Here $J$ and $U$ are the tunnelling and interaction parameters of the Hubbard model \cite{jaksch:1998}, $<>$ indicates a sum over nearest-neighbor lattice sites, $\mu$ is the chemical potential, $d$ is the spacing between lattice sites, $\hat{a}_{i,\alpha}$ is the operator in the Wannier basis that creates an $\alpha$ particle on lattice site $i$ ($\hat{n}_{i,\alpha}=\hat{a}_{i,\alpha}^\dag\hat{a}_{i,\alpha}$), $\hat{n}_{k,\beta}$ is the number operator for $k^{th}$ harmonic excitation of the parabolic trap, $\hat{n}_{\beta}(r)$ is the density operator for harmonic atoms at radius $r$ measured from the center of the parabolic potential, and $\phi_{i}(r)$ is the Wannier function centered at site $i$.  We have assumed that the $\alpha$ and $\beta$ atoms may experience different parabolic confining potentials with harmonic frequencies $\omega_\alpha$ and $\omega_\beta$.  Interactions between the two spin states are characterized by $\gamma_{ij} = 4\pi \hbar^2 a_{ij}/m$, where $a_{ij}$ is the s-wave scattering length between states $i$ and $j$.  In our physical implementation, $\alpha$ and $\beta$ are the $|1,-1\rangle$ and $|2,0\rangle$ states of $^{87}$Rb, so the masses $m$ are identical and the scattering length $a_{ij}$=98.1~$\pm$~0.1~$a_0$, where $a_0$ is the Bohr radius \cite{verhaar:2009,kokkelmanns:2009}. In collisions between atoms there is also the possibility of the atoms changing their spin projections (while conserving total spin) \cite{ho:1998,ohmi:1998}; we ignore these processes since they can be highly suppressed by applying a small magnetic field \cite{tojo:2009}.

\subsection{Effective Lattice for $\beta$ Atoms}

\begin{figure}
\begin{center}
\includegraphics[width=3in]{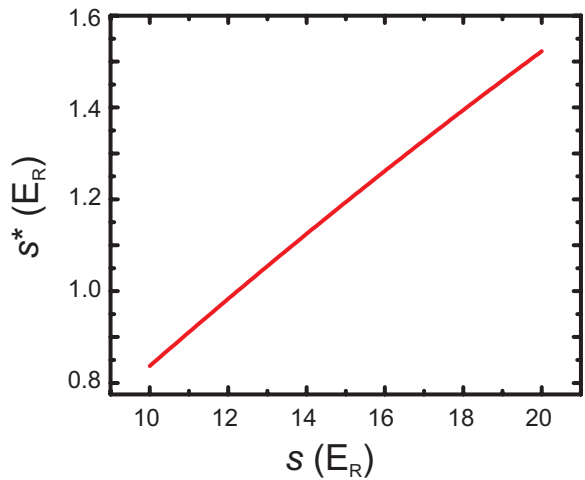}
\end{center}
\caption{\label{lattheight} Effective lattice potential depth $s^{*}$ for $\beta$ atoms arising from the interaction with a unit filled $\alpha$ lattice with potential depth $s$. We assume that the $\alpha$ atoms are confined in a retro-reflected 790~nm lattice.}
\end{figure}

One of the conditions for thermometry is that the impurity atoms have well-known thermodynamic properties. This may not be the case if they are strongly affected by the atoms in the lattice. We can estimate interaction effects by assuming that the $\alpha$ atoms are fixed in place and that inter-species interactions appear as a potential $\gamma_{\alpha \beta}  \sum_i \langle \hat{n}_{i,\alpha} \rangle \phi^2_{i}(r)$ for the $\beta$ atoms, where $\langle \rangle$ represents the expectation value.  For a sufficiently deep lattice, this will appear as an effective lattice for  the $\beta$ atoms with a potential depth of $\gamma_{\alpha \beta} \langle\hat{n}_{i,\alpha} \rangle  \phi^2_{0}(0)$. The effective lattice height is plotted as a function of the applied lattice depth in Figure \ref{lattheight} for $\langle\hat{n}_{i,\alpha} \rangle=1$.  These lattice heights are small even for strong $\alpha$ lattices and can be handled perturbatively using an effective mass formalism.

\subsection{Heat Capacity \label{sect:heatcap}}

Another condition for thermometry is that the impurity atoms do not change the temperature of the system being studied. However, as the lattice is applied the temperature of the lattice-bound atoms may change significantly \cite{blakie:2004}, and therefore heat must be transferred to/from the lattice atoms to maintain inter-species thermal equilibrium. If the impurity atoms start at $T_i$ and the lattice atoms at $T_f$, then the deviation of the final system temperature from $T_f$ due to thermalization is
\begin{equation}
\Delta T \approx \frac{N_{\beta}}{N_{\alpha}} \frac{ \int_{T_f}^{T_i} C_{\beta}(T^{\prime}) dT^{\prime}}{C_{\alpha}(T_f)}, \label{eqn:tchange}
\end{equation}
where $C_{\alpha},C_{\beta}$ are the per particle heat capacities of the two gases and we have assumed that $\Delta T$ is small. The desired bound on $\Delta T$ sets an upper bound on the size of the impurity system, $N_{\beta}/N_{\alpha}$.  Practically, this bound must be finite---the impurity cannot be set to arbitrarily low density given finite signal-to-noise ratio for time-of-flight imaging.\\

The heat capacity is defined as
\begin{equation}
C = \left. \frac{\partial \langle E \rangle}{\partial T} \right|_{N},
\end{equation}
where we assume the confining potential $U(r)$ is kept constant. For a harmonically trapped gas far from degeneracy the heat capacity is $3 N k_B$, while for a non-interacting Bose gas below $T_C$ the heat capacity is,
\begin{equation}
C = 10.8 N k_B \left(\frac{T}{T_C}\right)^{3}.
\end{equation}
Because of degeneracy, a harmonically trapped thermal gas may have a much larger heat capacity than a degenerate gas of interest. \\

To estimate bounds on the impurity atom number, we first consider a non-interacting gas in a combined lattice-parabolic potential \cite{rey:2005,mckay:2009,blakie:2004}. We assume that a gas of  $150,000$ atoms is prepared in a 50~Hz trap with a 70\% condensate fraction ($T=80$~nK), and that a 3D lattice is adiabatically turned on to $s=6$~$E_R$ with $J=0.051$~$E_R$; the harmonic confining frequency is kept constant. The temperature and condensate fraction in the lattice are $T^{\prime}=56.9$~nK and 83\% respectively, which are determined by calculating the entropy and fugacity $z$ in the lattice semi-classically from the grand canonical potential,
\begin{equation}
\Omega = -\left(\frac{2\pi k_B T}{d^2 m \omega^2}\right)^{3/2} k_B T \sum_{n=1}^{\infty} \frac{z^n e^{-6Jn/k_BT}}{n^{5/2}}I_0^{3}(2nJ/k_BT)
\end{equation}
where $d=790/2$~nm is the lattice spacing, and $I_0(x)$ is the modified Bessel function of the first kind. The  heat capacity in the lattice is 2.35~$k_B$ per particle, calculated according to $C = \left. \frac{\partial \langle E \rangle}{\partial T} \right|_N$. The gas must therefore absorb heat from the impurity in order to reach thermal equilibrium. Using (\ref{eqn:tchange}), we determine that the impurity state must consist of less than $14,500$ atoms to result in less than a 5\% change in $T^{\prime}$.\\

Interaction effects tend to reduce the heat capacity of atoms in a lattice, and therefore reduce the limit on the number of impurity atoms.  To estimate the impact of interaction effects, we use site-decoupled mean field theory \cite{sheshadri:1993,yoshimura:2008,spielman:2008} and the local density approximation to calculate the heat capacity. For the same initial conditions, but turning on the lattice to $s=17$ (keeping the parabolic potential fixed), $T^{\prime}=17$~nK and the heat capacity is 0.8~$k_B$ per particle in the lattice.  In this regime, 91\% of the atoms are in the Mott-insulator phase (at $T=0$), which significantly reduces the overall heat capacity.  For these conditions, the the impurity must be comprised of less than 500 atoms in order to limit the change in T' to 5\%.

\subsection{Thermalization}

The final practical constraint on this type of thermometry is sufficient thermal contact between the impurity and lattice-bound atoms. Adequate thermal contact is achieved when the thermalization rate, which is the rate for energy to be exchanged between spin states, is higher than atom loss and heating rates. For harmonically trapped atoms, thermalization has been extensively studied in the context of evaporative cooling, and the thermalization rate is proportional to the collision rate. For example, 2.5 s-wave collisions per atom are required for cross-dimensional thermalization in a trapped gas \cite{monroe:1993,snoke:1989}.   The total collision rate between species $i$ and $j$ is given by \cite{holland:2000},
\begin{eqnarray}
\gamma_{coll} & = & (1+\delta_{ij})4\pi a_{ij}^2 \overline{|v_i-v_j|} \int n_i(r) n_j(r) d^{3}r \label{eqn:collrate},
\end{eqnarray}
where $\overline{|v_i-v_j|}$ is the mean-relative-speed between species, and $n_i(r)$ and $n_j(r)$ are the atomic densities.\\

The general issue of thermalization in optical lattices is unresolved and is an active topic of current research \cite{griessner:2006,strohmaier:2009,eckstein:2009,kollath:2007}. Some insight into the problem may be gained from the literature on thermalization between species with different masses \cite{mosk:2001,delannoy:2001}, for which the thermalization rate is proportional to $\frac{4m_1 m_2}{(m_1+m_2)^2}$.  If we assume the effect of the lattice on thermalization is to change the effective mass of the lattice species, then the collision rate is reduced to approximately 90\% at $s=6$ and nearly 50\% at $s=10$ of the bare-mass value (using $m^{*}=\hbar^2/2d^2J$).\\

For comparing collision rates to loss and heating rates, we calculate the time between elastic collisions per $|2,0\rangle$ atom before turning on the lattice.  We use the parameters from Section \ref{sect:impurity}: $123,000$ atoms in the $|1,-1\rangle$ state, and $12,000$ atoms in the $|2,0\rangle$ state at $T=73$~nK. The $|2,0\rangle$ atoms are in a thermal state, and the condensate fraction for the $|1,-1\rangle$ gas is 76\%. The calculated time $\tau_{coll}=1/\gamma_{coll}$ for elastic collisions between $|2,0\rangle$ and $|1,-1\rangle$ atoms is 47~ms. Here we neglect degeneracy effects, and we assume zero velocity and a Thomas-Fermi density profile for the $|1,-1\rangle$ atoms.  This is the fastest elastic collision time in the system, as compared to collisions between $|2,0\rangle$ atoms ($\tau=300$~ms) and between $|2,0\rangle$ atoms and the $|1,-1\rangle$ thermal atoms ($\tau=225$~ms). We assume that this is the lowest relevant collision rate since turning on the lattice increases the density of the $|1,-1\rangle$ atoms.\\

Thermalization must compete with heating and loss processes, such as collisions with residual gas atoms (typically $\tau>100$~s), and three-body recombination and hyperfine relaxation in binary collisions involving $|2,0\rangle$ atoms. For pure condensates in $|1,-1\rangle$, three-body recombination is the limiting process with approximately a 30~s lifetime for the parameters considered in this work \cite{burt:1997}. Atoms in the $|2,0\rangle$ state involved in collisions can relax to the $F=1$ state, and convert the hyperfine energy ($E_{hf}/k_B \approx 0.3$~$K$) into kinetic energy. For collisions between $|2,0\rangle$ thermal atoms, we estimate a 12 s lifetime using the rate (measured for condensate atoms) from \cite{tojo:2009}. The most dominant loss process arises from collisions between $|2,0\rangle$ atoms and the $|1,-1\rangle$ condensate, which gives a lifetime of $\approx 830$~ms as estimated from the loss rate measured between a $|1,-1\rangle$ and $|2,1\rangle$ condensate in \cite{mertes:2007}. This rate is consistent with the negligible loss observed in Section \ref{sect:impurity} over 100~ms. While these rates may change with the lattice present, they appear to be sufficiently long such that heating in the lattice from spontaneous scattering---as discussed in Section \ref{sect:theory1}---will be the dominant process competing with thermalization.

\subsection{Limitations on Measuring the Impurity Temperature}

The proposed thermometry method depends on reliably measuring the temperature of the harmonically trapped gas, which is typically carried out by determining the expansion velocity after release from the trap. Given that we wish to avoid Bose condensation of the thermometer gas (in order to minimize interaction effects), we must therefore work at temperatures higher than $T_C=0.94\hbar \bar{\omega} N^{1/3}$. To maximize the dynamic range in temperature, both the number of atoms $N$ and the harmonic oscillator frequency $\omega$ can be decreased.  We note that N must already be quite small to minimize heat capacity effects, as discussed in Section \ref{sect:heatcap}. The lower bound on number and trap strength is ultimately set by technical issues, such as signal-to-noise in imaging. Reasonable lower bounds are $N=1000,\bar{\omega} = 2\pi (20Hz)$, for which $T_C=9$nK. This should be compared to the ``melting'' temperature of the Mott-Insulator \cite{gerbier:2007}, $T^{*}\approx 0.2U/k_B$, which is 15nK for $\lambda=790$nm and $s=16$. Therefore, this method should be able to probe temperatures in the Mott Insulator regime. It may be more difficult to reach the regime in which the layers between insulating regions are superfluid, which occurs below $T\approx zJ$, where $z$ is the coordination number. Since $z=6$ for a 3D cubic lattice, this gives a temperature of 6nK for $\lambda=790$nm and $s=16$. 

\section{\label{sect:impurity} Creating an Impurity}

A key component of our thermometry scheme is deterministically preparing an impurity state that is thermalized with the atoms of interest.  For the proof-of-principle experiments we discuss here, the $|F=2,m_F=0\rangle$ state acts as the thermometer for $|1,-1\rangle$ atoms confined in a lin-perp-lin lattice.  In this section, we show that adiabatic rapid passage driven by a microwave frequency magnetic field can be used to create a condensed mixture of the $|2,0\rangle$ and $|1,-1\rangle$ states that is out of thermal equilibrium.  The $|2,0\rangle$ condensate subsequently decays into a thermalized, low density component.\\

We create BECs of $^{87}$Rb in the $|1,-1\rangle$ state---details can be found in \cite{mckay:2009}, with several changes discussed here. Previously we worked with spin-polarized atoms in the $|1,-1\rangle$ state confined in a hybrid magneto-optical trap formed from a single-beam 1064~nm dipole trap and a quadrupole magnetic field. This method is not appropriate for a gas composed of multiple spin states with different values of $g_F m_F$. We now prepare a cold, but uncondensed, gas of $|1,-1\rangle$ atoms in the hybrid trap created using crossed 1064 nm dipole traps. We evaporatively cool the gas close to degeneracy in this trap, then simultaneously ramp the magnetic quadrupole off, a 3 G bias magnetic field on, and the dipole power up (to compensate gravity).  We evaporatively cool in the purely optical trap to the desired condensate fraction and then create a spin mixture via adiabatic rapid passage; information regarding the microwave source can be found in \cite{white:2009}.  At the end of this procedure, the dipole trap laser power is approximately 2.5~W, and the confining harmonic frequencies are 88.2~$\pm$~0.8, 29.8~$\pm$~0.5, and 92.8~$\pm$~0.7~Hz.\\

\begin{figure}
\begin{center}
\includegraphics[width=5.5in]{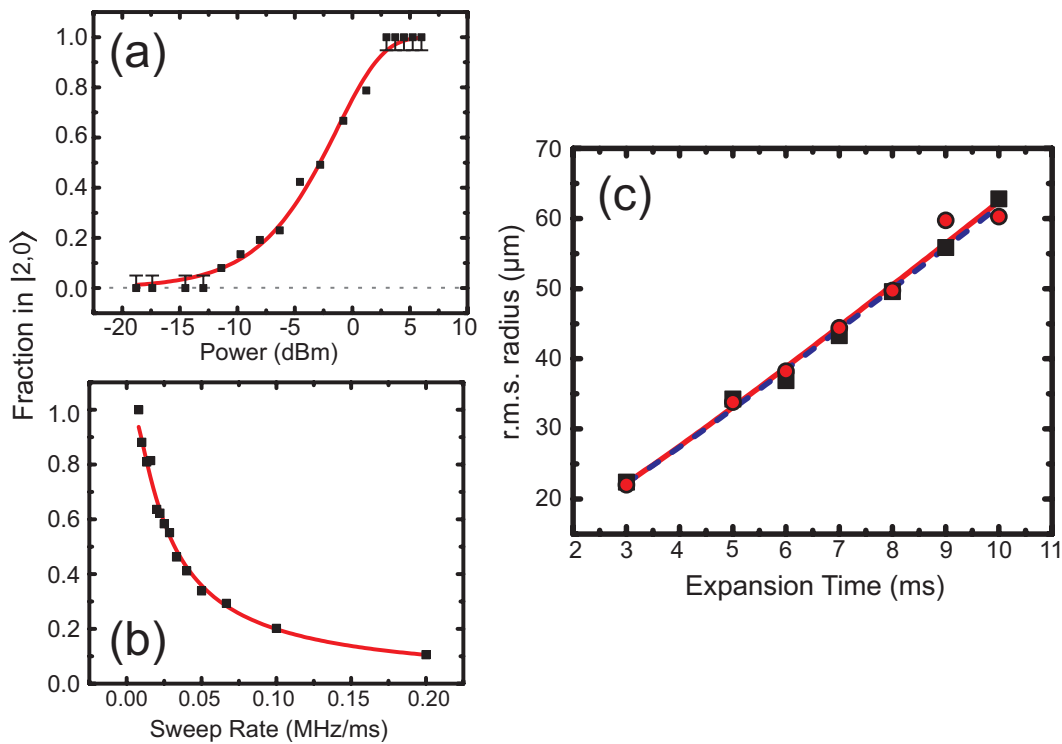}
\end{center}
\caption{Creating an impurity. Parts (a) and (b) show how the fraction of atoms in the $|2,0\rangle$ state can be controlled by varying the microwave power and sweep rate; the data in (a) were taken using a fixed 0.02 MHz/ms sweep rate.  Fits (red lines) are to the Landau-Zener theory for two states. The error bars in (a) represent a systematic uncertainty in determining atom number arising from finite signal-to-noise ratio in imaging.  Part (c) shows time-of-flight expansion data used to measure temperature for a 13\% $|2,0\rangle$ impurity $(\fullcircle)$ and $|1,-1\rangle$ $(\fullsquare)$ gas before transfer.  Images of the atom gas were fit to a Gaussian, and the fitted r.m.s. radius of the gas is shown vs. expansion time (red line $|2,0\rangle$, blue dashed line $|1,-1\rangle$).  The measured temperature was $393 \pm 20$~nK for the $|2,0\rangle$ gas and $383 \pm 13$~nK for the $|1,-1\rangle$ gas, thus illustrating that the microwave transfer preserves temperature.  \label{fig:createimpurity}}
\end{figure}

Figures \ref{fig:createimpurity}(a) and (b) show performance data for transferring atoms into the $|2,0\rangle$ state for a thermal gas.  For these data, a microwave sweep centered at 6832.65 MHz was applied to the atoms.  The number of atoms transferred into the $|2,0\rangle$ state was measured by hyperfine state selective imaging.  The fraction of atoms transferred into the $|2,0\rangle$ state can be smoothly varied either by adjusting the microwave power (Figure \ref{fig:createimpurity}(a)) or the sweep rate (Figure \ref{fig:createimpurity}(b)).  Time-of-flight expansion data shown in Figure \ref{fig:createimpurity}(c) demonstrates that microwave transfer does not affect the temperature of the gas.\\

For thermometry in the condensed regime, a mixture of $|1,-1\rangle$ and thermal $| 2,0\rangle$ atoms must be prepared so that the atoms used for thermometry are weakly interacting.  Na\"{i}vely, one would expect that adiabatic rapid passage as employed here could only be used to create a spinor condensate composed of two spin states.  Spinor condensates have been extensively studied \cite{myatt:1997,hall:1998,stenger:1998,ho:1998,ohmi:1998,chang:2005,kronjager:2005} and are a vibrant area of current research \cite{vengalattore:2009,sadler:2006}. However, these experiments typically probe the zero temperature regime.\\

\begin{figure}
\begin{center}
\includegraphics[width=5in]{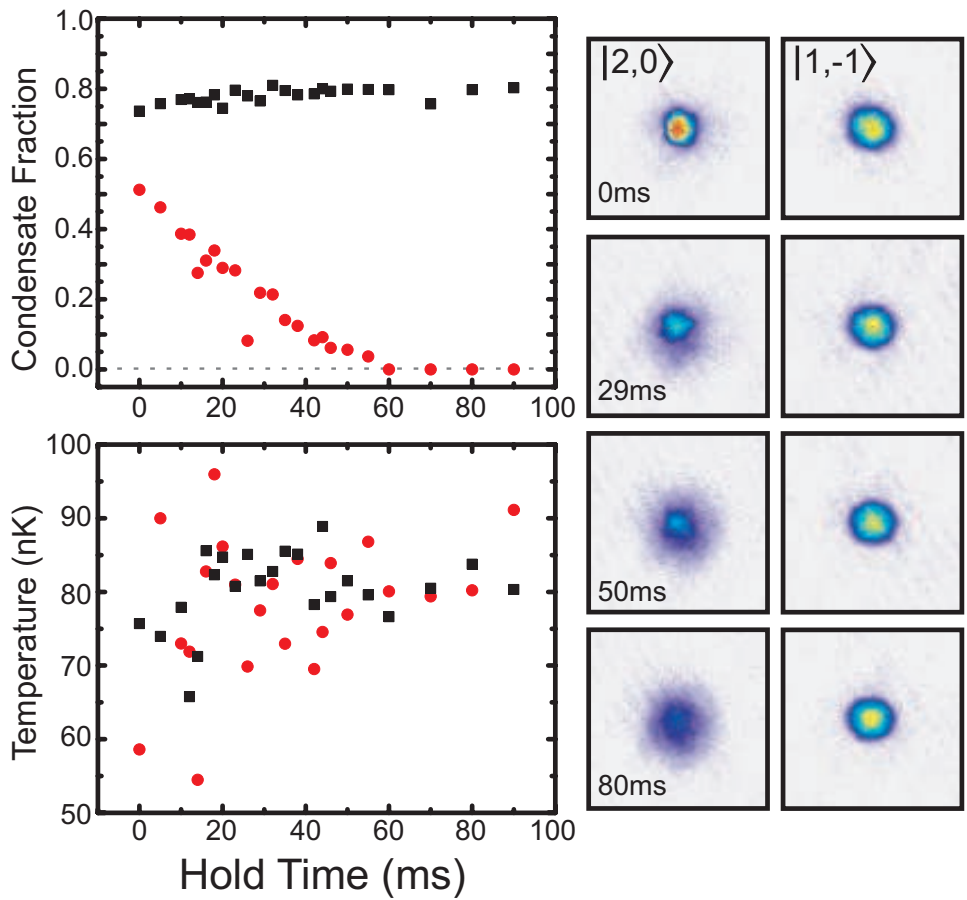}
\end{center}
\caption{Melting of an impurity condensate.  Condensate fraction and temperature of $|2,0\rangle$ ($\fullcircle$) and $|1,-1\rangle$ ($\fullsquare$) are shown after a microwave sweep that transfers  $9.0 \pm 0.5$\% of the atoms into the $|2,0\rangle$ state. Images of the $|2,0\rangle$ and $|1,-1\rangle$ gases are shown on the right for different hold times in the dipole trap after the transfer; approximately 10\% of the $|1,-1\rangle$ atoms are imaged using partial repumping \cite{mckay:2008}. There is a 5\% systematic error in both the temperature and condensate fraction due to uncertainties in the expansion time and the fitting procedure.  \label{fig:melting}}
\end{figure}

We find that, under the right conditions, creating a spin impurity can take the gas far from equilibrium into a state that decays into a mixture of $|1,-1\rangle$ condensed and $|2,0\rangle$ thermal components.  An example of this is shown in Figure \ref{fig:melting}.  Here we start with a $|1,-1\rangle$ condensate composed of $(135\pm8)\times10^3$ atoms, with a condensate fraction of $0.76 \pm 0.01$ at $73 \pm 5$ nK ($T_c$ is $144 \pm 3$ nK).  A swept microwave field transfers $9.0 \pm 0.5$\% of the atoms into the $|2,0\rangle$ state.  As this spin mixture is held in the dipole trap, the condensate fraction stays relatively constant for the $|1,-1\rangle$ atoms, yet decreases to zero for the $|2,0\rangle$ atoms in approximately 60 ms.  The simplest explanation of this phenomenon is a thermodynamic argument. The transfer is approximately isothermal, so the temperature of the $|2,0\rangle$ gas is unchanged. However, since $T_C \propto N^{1/3}$, the spin components have two different condensation temperatures: $140 \pm 3$ nK for the $|1,-1\rangle$ atoms (relatively unchanged by the microwave transfer) and $64 \pm 1$ nK for the $|2,0\rangle$ atoms.  As the system relaxes back to thermal equilibrium, the $|2,0\rangle$ condensate ``melts'', while the $|1,-1\rangle$ condensate remains unperturbed.  The decay timescale is roughly consistent with the elastic collision rate between $|2,0\rangle$ atoms and the $|1,-1\rangle$ condensate calculated in Section \ref{sect:theory2}.  After decay of the $|2,0\rangle$ condensate, the components are in thermal equilibrium, and the $|2,0\rangle$ component can be used for thermometry. Past work on spinor gases out of thermal equilibrium can be found in \cite{lewandowski:2003,erhard:2004,schmaljohann:2004}.\\

\section{\label{sect:1Dlattice} A 1D Spin-Dependent Lattice}

\begin{figure}
\begin{center}
\includegraphics[width=5.5in]{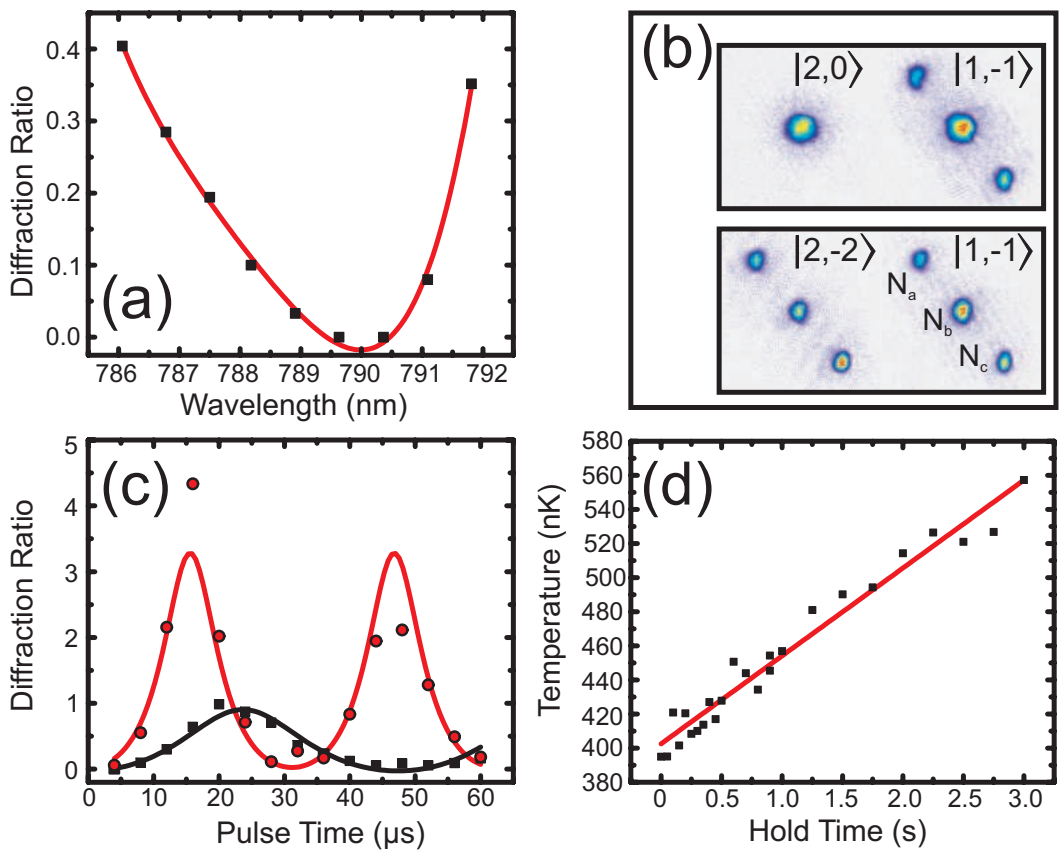}
\end{center}
\caption{Properties of a 1D spin-dependent lattice.  Part (a) shows the diffraction ratio $\frac{N_A+N_C}{N_B}$ (see (b)) for $|1,-1\rangle$ atoms from a 20~$\mu$s lin-lin lattice pulse as the laser wavelength is varied; the minimum near 790 nm is where the scalar AC Stark shift vanishes. The red line is present to guide the eye.  Images of diffracted atoms in different spin states are shown in (b); two separate images are taken with the same lattice pulse power and duration (20 $\mu$s).  Data taken to calibrate the lattice potential depth for the $|2,-2\rangle$ ($\fullcircle$) and $|1,-1\rangle$ ($\fullsquare$) states are shown in (c). The data are fit (solid lines) to a two-band model that determines an energy difference between ground and the second-excited bands of $32.0 \pm 0.5$~kHz for the $|2,-2\rangle$ atoms and $21.8 \pm 5$~kHz for the $|1,-1\rangle$ atoms.  This corresponds to lattice potential depths of $11.8 \pm 0.3$ $E_R$ and $6.6 \pm 0.3$~$E_R$, respectively. The ratio of the lattice potential depths is $1.8 \pm 0.1$, which is nearly consistent with the prediction from (\ref{eqn:statedeplattice2}). Part (d) shows the heating of $|1,-1\rangle$ atoms loaded into the forward beam of a 10~$E_R$ lattice (the retro-reflected beam is blocked). \label{fig:latsetup}}
\end{figure}

The other ingredient essential to the thermometry scheme is transferring the atoms into a spin-dependent lattice.  To create the lattice, we use a setup similar to that in \cite{mckay:2009}, except that we have added quarter-wave plates into the retro beam paths to rotate the laser polarization by $90^{\circ}$. We use 790~nm light to create the lattice for two reasons. First, as explained in Section \ref{sect:theory1}, employing 790~nm light optimizes the ratio of lattice depth to spontaneous scattering for a lin-perp-lin lattice.  Second, again from Section \ref{sect:theory1}, working at 790~nm minimizes problems introduced by laser polarization impurities---there is no scalar light shift at this wavelength since the AC Stark shift from the D1 and D2 transitions cancel. Small imperfections in laser polarization are magnified because the lattice potential arises from an interference effect.  For example, for lin-perp-lin lattice at 785~nm, it would take a 5\% impurity in the retro beam polarization to create a lattice for the $|2,0\rangle$ state with half the potential depth as for the $|1,-1\rangle$ state.  By using 790~nm light to make the lattice, polarization impurities do not contribute to a parasitic lattice, and at worst increase the heating rate. The absence of the scalar light shift at 790~nm is evident by observing diffraction of $|1,-1\rangle$ atoms from a pulsed 1D lin-lin lattice for different lattice wavelengths, as shown in Figure \ref{fig:latsetup}(a).\\

To verify the properties of the 790~nm spin-dependent lattice, we measured diffraction of the atoms from the lattice by transiently pulsing the lattice and then turning off the trap.  Figure \ref{fig:latsetup}(b) shows images of diffracted atoms in the $|1,-1\rangle$, $|2,-2\rangle$, and $|2,0\rangle$ states.  Figure \ref{fig:latsetup}(c) shows that the lattice potential depth---measured by a fit to the diffracted fraction as the pulse time is varied---scales as $g_F m_F$, as predicted by (\ref{eqn:statedeplattice2}).  Finally, we characterized heating due to the lattice light by turning on only the forward beam of a 10~$E_R$ lattice for a gas of thermal atoms. By holding the atoms for a variable amount of time, we observe a heating rate of $52 \pm 2$~nK/sec, which corresponds to a heating power of $0.88 \pm 0.03E_R/\mathrm{sec}$. This roughly matches the predicted power of approximately $0.6 E_R/\mathrm{sec}$, which assumes equal power in the forward and retro beams. \\

\begin{figure}
\begin{center}
\includegraphics[width=4in]{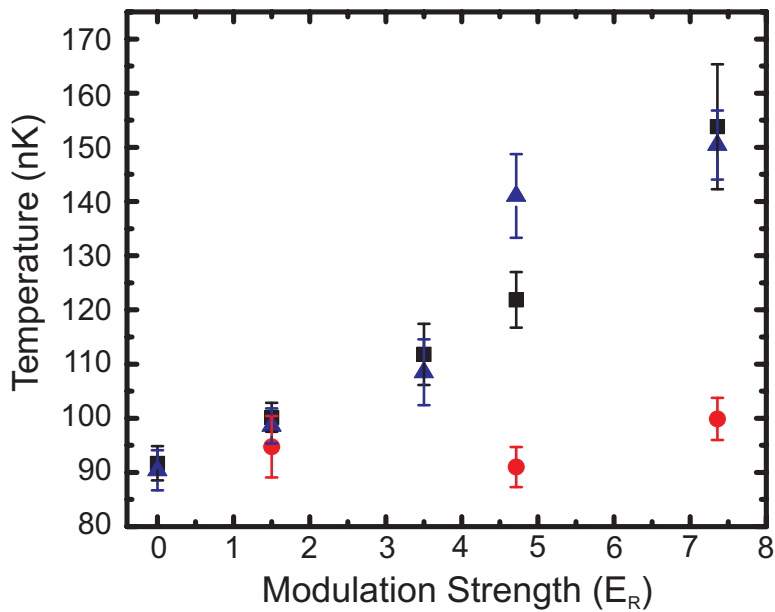}
\end{center}
\caption{Temperature of a 10\%/90\% mixture of $|2,0\rangle$ ($\blacktriangle$)/$|1,-1\rangle$ ($\blacksquare$) atoms and a spin-polarized $|2,0\rangle$ ($\fullcircle$) gas after modulating the lattice potential.  The error bars represent the uncertainty from the fit to time-of-flight expansion data used to determine temperature. \label{fig:shaking}}
\end{figure}

To demonstrate the main principle behind spin-dependent thermometry---that atoms in the $|2,0\rangle$ state can be sensitive to the temperature of the $|1,-1\rangle$ atoms without being bound to the lattice potential---we use the lattice to heat $|2,0\rangle$ atoms through their thermal contact with $|1,-1\rangle$ atoms. To heat the atoms, the lattice potential depth is modulated at 24~kHz, which is near the frequency separating the ground and second excited bands of the lattice (28.4~kHz). We compare the resulting temperature of the gas when there is a 10\% impurity of $|2,0\rangle$ atoms to when all of the atoms have been transferred into the $|2,0\rangle$ state. The lattice is turned on in 20~ms to 10~$E_R$ (calibrated using the $|1,-1\rangle$ atoms), the amplitude is modulated for 6 ms at 24~kHz, the lattice turned off in 10~ms, and the atoms are permitted to thermalize for 220~ms. The temperature of the two components is measured simultaneously via time-of-flight expansion velocity using a magnetic field gradient to separate the spin states. The measured temperature for different modulation amplitudes is shown in Figure \ref{fig:shaking}. The $|2,0\rangle$ atoms are heated only when the $|1,-1\rangle$ atoms are present. This is the first step towards thermometry using the $|2,0\rangle$ atoms, and a useful technique for selectively heating or exciting motion for one state.

\section{\label{sect:3dlattice} 3D Lattice}

In this section, we present the first demonstration of atoms trapped in a 3D spin-dependent lattice in the strongly-correlated regime.  A requirement for creating a 3D spin-dependent lattice is that $\hat{k}\cdot \hat{B} \ne 0$ for all lattice wavevectors. Because of this condition $\vec{B}$ cannot point directly along any of the wavevectors, as was the case in previous experiments with spin-dependent lattices \cite{mandel:2003,mandel:2003b,karski:2009}. In our experiment we satisfy this condition using the geometry
\begin{equation}
\fl \hat{k}_1 = \frac{1}{2}\left(\begin{array}{c}1\\-\sqrt{2}\\-1\end{array}\right), \hat{k}_2 = \frac{1}{2}\left(\begin{array}{l}1\\\sqrt{2}\\-1\end{array}\right), \hat{k}_3 = -\frac{1}{\sqrt{2}}\left(\begin{array}{c}1\\0\\1\end{array}\right), \hat{B} = \left(\begin{array}{c}1\\0\\0\end{array}\right),
\end{equation}
where $\hat{z}$ is opposite to gravity and we image along $\hat{y}$.  In this configuration, $\hat{k}\cdot\hat{B}$ is 1/2 for two of the beams and $1/\sqrt{2}$ for the other, so that per unit intensity the lattice along one direction is 40\% larger. This configuration is fairly close to the ideal case, for which $\hat{k}\cdot\hat{B}=1/\sqrt{3} \approx 0.58$.\\

\begin{figure}
\begin{center}
\includegraphics[width=3in]{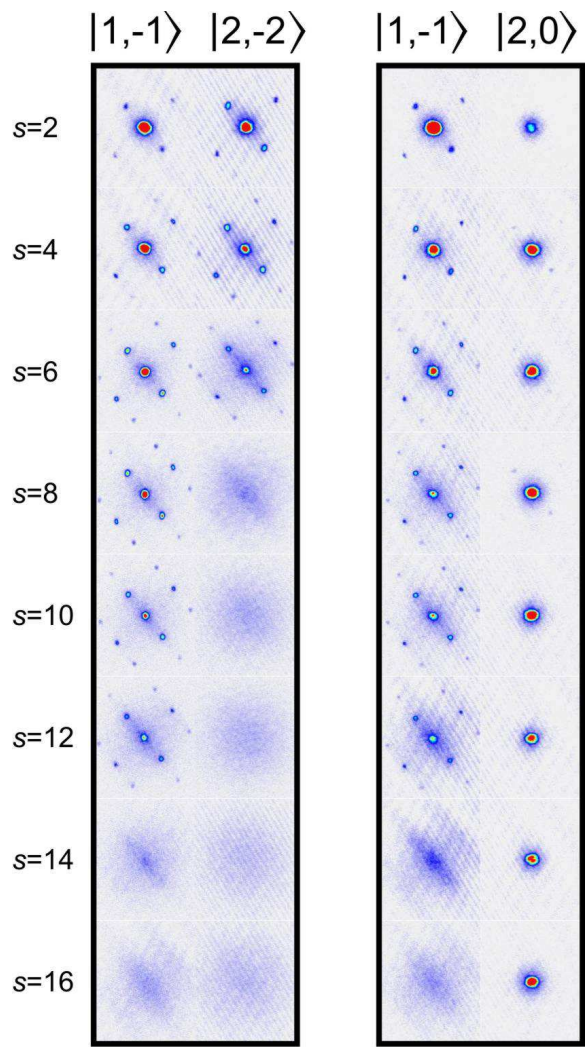}
\end{center}
\caption{Time-of-flight images for two different spin mixtures transferred into a 3D lin-perp-lin lattice. The lattice potential depth for each set of images is given on the left, calibrated for the $|1,-1\rangle$ atoms.  The lattice is turned on using an exponential ramp in 50 ms. Images were taken using a single shot, and the spin states were separated using a magnetic field gradient during expansion. Empty space between the different spin states in the images has been cropped out.\label{fig:3Dlattice}}
\end{figure}

Using this arrangement of lattice beams, we load two different spin mixtures into the lattice: an approximately equal mixture of $|2,-2\rangle$ and $|1,-1\rangle$ atoms, and an approximately equal mixture of $|2,0\rangle$ and $|1,-1\rangle$ atoms. This former is an interesting state because it has imposed anti-ferromagnetic spin ordering due to the potential.  Density profiles measured after suddenly shutting off the lattice and 15 ms of expansion are shown in Figure \ref{fig:3Dlattice}. The different spin states are separated by a magnetic field gradient, and images are shown for different lattice potential depths calibrated for the $|1,-1\rangle$ atoms. \\

For a mixture of $|2,-2\rangle$ and $|1,-1\rangle$ atoms, the time-of-flight images for both components are consistent with the transition from a superfluid to a Mott-insulator \cite{greiner:2002} state, which is predicted from mean field theory to occur at $s=12.7$~$E_R$.  Because the $|2,-2\rangle$ atoms experience twice the lattice potential depth at the same intensity, this transition occurs at half of the lattice intensity as for the $|1,-1\rangle$ atoms.  If we slowly turn off the lattice in 10 ms from $s=16$ (for the $|1,-1>$) atoms, we recover approximately 10\% condensate fraction for each component, indicating that the transfer into the lattice was partially reversible.\\

For a mixture of $|2,0\rangle$ and $|1,-1\rangle$ atoms, the time-of-flight images are consistent with the $|1,-1\rangle$ atoms experiencing a lattice and the $|2,0\rangle$ atoms only being trapped harmonically. Extremely weak diffraction features can be seen in several of the $|2,0\rangle$ images, which may be due to interactions with the $|1,-1\rangle$ atoms, as discussed in Section \ref{sect:theory2}. We checked that these features are not directly the result of the lattice potential in 1D. Recent results \cite{pertot:2009} suggest that these features may be due to interactions between $|2,0\rangle$ and $|1,-1\rangle$ atoms during time-of-flight. Unlike the $|1,-1\rangle$ and $|2,-2\rangle$ mixture, which are nearly physically separated since the lattice potentials are $180^{\circ}$ out of phase, the $|2,0\rangle$ atoms are free to interact with the $|1,-1\rangle$ atoms. Although a more careful study is needed, there is no obvious change in the $|1,-1\rangle$ images induced by the presence of the $|2,0\rangle$ atoms. 

\section{\label{sect:conc} Conclusions}

Spin-dependent lattices are a promising system for thermometry of strongly correlated phases.  We have demonstrated proof-of-principle experiments for thermometry in which a weakly interacting gas of atoms in a state with $g_F m_F=0$ can be used to determine the temperature of a strongly correlated, lattice-bound gas.   We have also shown that spin-dependent lattices can be realized in 3D and in the strongly correlated limit.  Thermometry using $g_F m_F=0$ atoms in such a system remains to be demonstrated. Several experimental and theoretical challenges are also unresolved, such as accurately measuring the effect of heating in the lattice, measuring and calculating thermalization rates, and determining the minimal impurity size required for accurate time-of-flight thermometry. Demonstration of this technique will be an important step towards quantum simulation in lattices, and will complement other thermometry techniques and ongoing efforts to cool atomic gases to ever lower temperatures in a lattice.

\ack

This work was done at the University of Illinois, supported by the DARPA OLE program and the National Science Foundation (award 0448354). D. McKay acknowledges support from NSERC. D. McKay would also like to thank the Thywissen lab at the University of Toronto for hospitality during the preparation of this manuscript. We thank Josh Zirbel for careful reading of this manuscript. 

\section*{References}

\bibliography{spindep_paper}
\bibliographystyle{unsrt}

\end{document}